\documentclass[prl,twocolumn,showpacs,preprintnumbers,amsmath,amssymb,floatfix,showkeys,twoside]{revtex4}

\usepackage{graphicx}
\usepackage{dcolumn}
\usepackage{bm}

\usepackage{epsfig,latexsym}

\begin{document}

\def\bef{\begin{figure}}
\def\eef{\end{figure}}
\newcommand{\ans}{ansatz }
\newcommand{\be}[1]{\begin{equation}\label{#1}}
\newcommand{\beq}{\begin{equation}}
\newcommand{\ee}{\end{equation}}
\newcommand{\beqn}[1]{\begin{eqnarray}\label{#1}}
\newcommand{\eeqn}{\end{eqnarray}}
\newcommand{\bd}{\begin{displaymath}}
\newcommand{\ed}{\end{displaymath}}
\newcommand{\dub}[2]{\left(\begin{array}{c}{#1}\\{#2}
\end{array}\right)}
\newcommand{\dubs}[2]{\begin{array}{l}{#1}\\{#2} \end{array}}
\newcommand{\mat}[4]{\left(\begin{array}{cc}{#1}&{#2}\\{#3}&{#4}
\end{array}\right)}
\newcommand{\matr}[9]{\left(\begin{array}{ccc}{#1}&{#2}&{#3}\\
{#4}&{#5}&{#6}\\{#7}&{#8}&{#9}\end{array}\right)}
\newcommand{\matrr}[6]{\left(\begin{array}{cc}{#1}&{#2}\\
{#3}&{#4}\\{#5}&{#6}\end{array}\right)}
\newcommand{\cvb}[3]{#1^{#2}_{#3}}
\def\lsim{\raise0.3ex\hbox{$\;<$\kern-0.75em\raise-1.1ex
e\hbox{$\sim\;$}}}
\def\gsim{\raise0.3ex\hbox{$\;>$\kern-0.75em\raise-1.1ex
\hbox{$\sim\;$}}}
\def\abs#1{\left| #1\right|}
\def\simlt{\mathrel{\lower2.5pt\vbox{\lineskip=0pt\baselineskip=0pt
           \hbox{$<$}\hbox{$\sim$}}}}
\def\simgt{\mathrel{\lower2.5pt\vbox{\lineskip=0pt\baselineskip=0pt
           \hbox{$>$}\hbox{$\sim$}}}}
\def\unity{{\hbox{1\kern-.8mm l}}}
\newcommand{\eps}{\varepsilon}
\def\ep{\epsilon}
\def\ga{\gamma}
\def\Ga{\Gamma}
\def\om{\omega}
\def\OM{\Omega}
\def\la{\lambda}
\def\La{\Lambda}
\def\al{\alpha}
\newcommand{\ov}{\overline}
\renewcommand{\to}{\rightarrow}
\renewcommand{\vec}[1]{\mbox{\boldmath$#1$}}
\def\tm{{\widetilde{m}}}
\def\mcirc{{\stackrel{o}{m}}}
\newcommand{\dm}{\delta m}
\newcommand{\tanb}{\tan\beta}
\newcommand{\nbar}{\tilde{n}}

%

\newcommand{\Dsusy}{{susy \hspace{-9.4pt} \slash}\;}
\newcommand{\DCP}{{CP \hspace{-7.4pt} \slash}\;}
\newcommand{\mc}{\mathcal}
\newcommand{\gr}{\mathbf}
\renewcommand{\to}{\rightarrow}
\newcommand{\gtc}{\mathfrak}  
\newcommand{\wh}{\widehat}
\newcommand{\br}{\langle}
\newcommand{\kt}{\rangle}


\def\lsim{\mathrel{\mathop  {\hbox{\lower0.5ex\hbox{$\sim$}
\kern-0.8em\lower-0.7ex\hbox{$<$}}}}}
\def\gsim{\mathrel{\mathop  {\hbox{\lower0.5ex\hbox{$\sim$}
\kern-0.8em\lower-0.7ex\hbox{$>$}}}}}

\def\nn{\\  \nonumber}
\def\de{\partial}
\def\brf{{\mathbf f}}
\def\bbf{\bar{\bf f}}
\def\bF{{\bf F}}
\def\bbF{\bar{\bf F}}
\def\bA{{\mathbf A}}
\def\bB{{\mathbf B}}
\def\bG{{\mathbf G}}
\def\bI{{\mathbf I}}
\def\bM{{\mathbf M}}
\def\bY{{\mathbf Y}}
\def\bX{{\mathbf X}}
\def\bS{{\mathbf S}}
\def\bb{{\mathbf b}}
\def\bh{{\mathbf h}}
\def\bg{{\mathbf g}}
\def\bla{{\mathbf \la}}  
\def\bmu{\mathbf m }
\def\by{{\mathbf y}}
\def\bmu{\mbox{\boldmath $\mu$} }
\def\bunity{{\mathbf 1}}
\def\cA{{\cal A}}
\def\cB{{\cal B}}
\def\cC{{\cal C}}
\def\cD{{\cal D}}
\def\cF{{\cal F}}
\def\cG{{\cal G}}
\def\cH{{\cal H}}
\def\cI{{\cal I}}
\def\cL{{\cal L}}
\def\cM{{\cal M}}
\def\cN{{\cal N}}
\def\cO{{\cal O}}
\def\cR{{\cal R}}
\def\cS{{\cal S}}   
\def\cT{{\cal T}}   
%

\newcommand{\mpl}{M_{\mathrm{Pl}}}




\title{Neutron -- Mirror Neutron Oscillations: How Fast Might They Be?}

\author{Zurab Berezhiani$^{a, }$}
\email{zurab.berezhiani@aquila.infn.it}

\author{Lu{\'{\i}}s Bento$^{b, }$}
\email{lbento@cii.fc.ul.pt}

\affiliation{ \vspace{5pt}
$^a$Dipartimento di Fisica, Universit\`a di L'Aquila, 
I-67010 Coppito, AQ, Italy \\ 
 and Laboratori Nazionali del Gran Sasso, INFN,
 I-67010 Assergi, AQ, Italy \\
$^b$Faculdade de Ci\^ encias, 
Centro de F\'{\i}sica Nuclear da Universidade de Lisboa, 
Universidade de Lisboa, \\
Avenida Professor Gama Pinto 2, 1649-003 Lisboa, Portugal
}


\begin{abstract}
We discuss the phenomenological implications of the neutron ($n$) oscillation  
into the mirror neutron ($n'$), a hypothetical particle 
exactly degenerate in mass with the neutron but sterile to normal matter. 
We show that the present experimental data allow a maximal $n-n'$ oscillation  
in vacuum with a characteristic time $\tau$ much shorter than the neutron 
lifetime, in fact as  small as $1$ sec. 
This phenomenon may manifest in neutron disappearance and regeneration
experiments perfectly  accessible to present experimental capabilities and 
may also have interesting astrophysical consequences, 
in particular for the propagation of ultra high energy cosmic rays. 
\end{abstract}

\pacs{14.20.Dh, 11.30.Fs, 12.60-i, 98.70.Sa}
\keywords{neutron oscillation, mirror world, cosmic rays }

\maketitle



The idea that there may exist a mirror world, a hidden parallel sector 
of particles that is an exact duplicate of our observable world, 
has attracted a significant interest over the  
years \cite{LY,Holdom,FV,ortho,BDM,PLB98,BB-PRL,BCV} 
(for reviews, see \cite{IJMPA-B,IJMPA-F}). 
Such a theory is based on the product $G\times G'$ of two identical gauge 
factors with identical particle contents, 
where ordinary (O) particles belonging to $G$ are singlets of $G'$, 
and mirror (M) particles belonging to $G'$ are singlets of $G$.  
Mirror parity under the proper interchange of  $G\leftrightarrow G'$ and 
the respective matter fields makes the Lagrangians of both sectors identical
to each other. 
Such a situation can emerge e.g.\ in the context of $E_8\times E'_8$ 
superstring theory.

Under this hypothesis, the Universe should contain along with the 
ordinary photons, electrons, nucleons etc., also their mirror partners 
with exactly the same masses.  
Invisible M-matter, interacting with O-matter via gravity, could 
be a viable dark matter candidate~\cite{BB-PRL,BCV,IJMPA-B,IJMPA-F}.

Besides gravity, the two sectors could communicate by other means.
In particular, any neutral O-particle, elementary or composite,   
can have a mixing with its M-twin. 
For example, kinetic mixing between ordinary and mirror photons \cite{Holdom} 
can be revealed in the ortho-positronium oscillation \cite{ortho}  
and can be tested also with dark matter detectors \cite{IJMPA-F}. 
Ordinary neutrinos can mix with mirror neutrinos and 
thus oscillate into these sterile species \cite{FV},
as well as neutral pions and Kaons into their mirror partners.    
Such mixings may be induced by interactions 
between the O- and M-fields mediated by messengers like 
pure gauge singlets or 
extra gauge bosons acting with both sectors~\cite{PLB98,IJMPA-B}. 

In this letter we explore the mixing between the ordinary neutron $n$ 
and mirror neutron $n'$ due to a small  mass term 
$\dm \,(\ov{n} n' + \ov{n}' n)$. 
We show that the existing experimental data do not exclude a rather fast 
$n -n'$ oscillation, with a timescale $\tau =\dm^{-1} \sim 1$ s. 
Such an intriguing possibility can be tested in small scale "table-top" 
experiments, and it can have strong astrophysical implications,
in particular for ultra high energy cosmic rays. 


Let us take the minimal gauge symmetry $G = SU(3) \times SU(2) \times U(1)$ 
for the O-sector that contains the Higgs doublet $\phi$, 
and quarks and leptons:  
left $q_L = (u, d)_L $,  $l_L = (\nu, e)_L$, and right $u_R, d_R,  e_R$    
\cite{footnote1}.
As usual, we assign a global lepton charge $L = 1$ to leptons and 
a baryon charge $B = 1/3$ to quarks. If $L$ and $B$ were exactly 
conserved then the Majorana masses of neutrinos would be forbidden 
and the proton would be stable.

However, $L$ and $B$ are not perfect quantum numbers.
They are related to accidental global symmetries possessed by the 
Standard Model Lagrangian at the level of  renormalizable couplings, 
which can be however explicitly broken by higher order operators 
cutoff by large mass scales $M$. 
In particular, the well-known 
D=5 operator 
$\cO_5 \sim(l\phi)^2 /M$ ($\Delta L=2$),
yields, after inserting the Higgs 
vacuum expectation value (VEV) $\langle \phi\rangle$,
small Majorana masses for neutrinos, 
$m_\nu \sim \langle \phi\rangle^2/M$  \cite{Weinberg}, 
while the D=9 operators 
$\cO_9 \sim (udd)^2 /M^5$ or $(qqd)^2 /M^5$  ($\Delta B=2$) 
lead to neutron - antineutron
($n - \nbar$) oscillation phenomenon~\cite{nnbar}.

Suppose now that there exists a hidden M-sector 
with a gauge symmetry $G' = SU(3)' \times  SU(2)'\times U(1)'$,
a mirror Higgs doublet $\phi'$, 
and mirror quarks and leptons:  
$q'_L = (u', d')_L$, $l'_L = (\nu', e')_L$ and  $u'_R, d'_R, e'_R$, 
where one assigns a lepton charge $L'=1$ to mirror leptons 
and a baryon charge $B'=1/3$ to mirror quarks.
Mirror parity $G\leftrightarrow G'$ tells that  
all coupling constants (gauge, Yukawa, Higgs) 
are identical for both sectors, 
the O- and M-Higgses have equal VEVs, 
$\langle \phi\rangle = \langle \phi'\rangle$,  
and hence the mass spectrum of mirror particles 
is exactly the same as that of ordinary ones.  
In addition, if the O-sector contains $L$ and $B$ violating 
operators like $\cO_5$ and $\cO_9$, 
then analogous $L'$ and $B'$ violating operators $\cO'_5$ and $\cO'_9$ 
should be  included in the M-sector. 

Moreover, there can exist higher order operators that 
couple gauge singlet combinations of O- and M-particles. 
In particular, the D=5 operator 
\be{l-lpr}
\cO^{\rm mix}_5 \sim  \frac{1}{M} (l \phi)(l' \phi')  
\ee
induces the mixing between the ordinary and mirror neutrinos \cite{FV},  
and the D=9 operators 
coupling three ordinary and three mirror quarks,
\be{n-npr}
\cO_9^{\rm mix} \sim 
\frac{1}{\cM^5} (udd)(u'd'd')  +  \frac{1}{\cM^5} (qqd)(q'q'd') \; , 
\ee
result in ordinary neutron - mirror neutron mixing.  
On the other hand, if one postulates conservation of 
the combined quantum number $\bar{B}= B - B'$,  
the operators $\cO_9, \cO'_9$ are forbidden 
while the operator  $\cO^{\rm mix}_9$ is allowed. 
Hence, the exact $\bar{B}$  conservation would suppress 
the $n-\nbar$ oscillation but not the $n-n'$ oscillation~\cite{footnote2}.
Taking into account that the matrix elements of the operators 
(\ref{n-npr}) between the neutron states are typically 
$\sim 10^{-4}$ GeV$^6$,
one estimates the mass mixing term
between $n$ and $n'$ states as: 
\be{deltam}
\dm \sim \left(\frac{10 \, {\rm TeV} }{\cM}\right)^5 \times 
10^{-15} \, {\rm eV}	\; .
\ee

One could naively think that the bound on $n -n'$ mixing is 
nearly as strong as the one on $n - \nbar$ mixing: 
$\dm_{n\nbar} < 10^{-23}$ eV,  or 
$\tau_{n\nbar}=\dm_{n\nbar}^{-1} > 10^8$ s,  
which follows from the direct experimental search of
free neutron oscillation into antineutron ($\nbar$) \cite{Grenoble},  
as well as from the limits on nuclear stability: 
a stable nucleus $(A,Z)$ would decay into states $(A-2,Z)$ or $(A-2,Z-1)$ 
due to the conversion $n\to \nbar$ followed by $\nbar n$ or $\nbar p$ 
annihilation into pions 
with a total energy of nearly two nucleon masses \cite{Chetyrkin}.
However, we show below that the bound on $n-n'$ oscillation is 
many orders of magnitude weaker,  $\tau =\dm^{-1} > 1$ s.
This indeed is quite surprising! The $n-n'$ oscillation time can be 
much smaller than the neutron lifetime $\tau_n \simeq 10^3$ s.

As far as mirror neutrons are invisible, the $n-n'$ oscillation can manifest 
experimentally only as a neutron disappearance. If $\tau \ll \tau_n$, 
strictly free neutrons would oscillate many times with a maximal mixing angle ($\theta=45^\circ$) before they decay. 
Thus, instead of the exponential law $P(t)=\exp(- t/\tau_n)$
for the neutron survival probability, one would observe
the oscillating behaviour $P(t)=\cos^2(t/\tau)\exp(-t/\tau_n)$.
Is not this immediately excluded by the experiments measuring
the neutron lifetime with great accuracy?  The answer is no! 
Simply because in these experiments neutrons and mirror neutrons 
are subject to very different conditions.

The evolution of free non-relativistic neutrons is described
by the effective Hamiltonian in $n-n'$ space,
\be{Hn-npr}
H = \mat{m -i\Gamma/2 - V}{\dm} {\dm} {m -i\Gamma/2  - V^\prime}  \;,
\ee
where $m$ is the neutron mass and $\Gamma=1/\tau_n$ its decay width,
which, due to exact mirror parity, are precisely the same for mirror neutrons.
However, the potentials $V$ and $V'$ felt by $n$ and $n'$ are not 
quite the same.
Namely, since the experiments are done on Earth,
the terrestrial magnetic field, $B\simeq 0.5$ G, induces an
effective contribution $V= \mu B\simeq 3\times 10^{-12}$ eV,
where  $\mu$ is the neutron magnetic moment. 
On the other hand, $V'=0$ as far as no mirror magnetic fields 
exist on Earth.
Thus, for $\dm \ll V$, the effective mixing angle between $n$ and $n'$ states  
is $\theta_{\rm eff} \approx \dm/V$ while the oscillation time is   
$\tau_{\rm eff}\approx 2/V \simeq 4 \times 10^{-4}$ s, 
and hence  the average transition probability becomes 
$\bar{P}_{nn'}\simeq 2\, (\dm/V)^2$. 
For example, for $\dm =10^{-15}$ eV one has 
$\bar{P}_{nn'}\approx 2\times 10^{-7}$,  
and such a small disappearance effect could hardly be observed 
in the experiments.  
Thus, to improve the experimental sensitivity 
the magnetic field should be suppressed. 

In the ILL-Grenoble experiment designed to search for
neutron - antineutron oscillation, the magnetic field 
was reduced to $B\sim 10^{-4}$ G \cite{Grenoble}. 
Cold neutrons propagated in vacuum with an average speed of $600$ m/s 
and effective time of flight $t \simeq 0.1$ s, 
in a $\mu$-metal vessel shielding the magnetic field.
No antineutrons were detected and the limit
$\tau_{n\nbar} > 0.86 \times 10^8$ s was reported.
Clearly, the search for $n-n'$ oscillation was not the aim of this experiment,  
but it can be used to set a crude limit on the timescale $\tau$. 
For $\tau$ larger than the neutron propagation time $t\simeq 0.1$ s, 
the oscillation probability is $P_{nn'}(t)\approx (t/\tau)^2$. 
By monitoring the neutron beam intensity it was observed that
about 5$\%$ of neutrons disappeared \cite{Grenoble}.
If this deficit could be entirely due to $n-n'$ oscillation, 
this would imply $\tau \approx 0.45$ s. 
However, as far as  most of the losses can be attributed to 
scatterings with the walls in the drift vessel, 
one can assume rather conservatively that no more than 
1$\%$ of losses were due to $n-n'$ oscillation, 
and thus obtain a bound $\tau  > 1$ s,  or $\dm < 10^{-15}$ eV.

Let us discuss whether the bounds from nuclear stability, 
which give the strongest limit on $\tau_{n\nbar}$,
are applicable also to the case of $n-n'$ oscillation.
One may naively think that it could destabilize nuclei as follows:  
in a stable nucleus $(A,Z)$ (e.g.\ $^{16}$O) $n$ oscillates into $n'$. 
Then $n'$, or its $\beta$-decay $n'\to p'e'\tilde{\nu}'_e$ products,  
can escape from the nucleus  thus producing an unstable isotope $(A-1,Z)$  
($^{15}$O) 
whose characteristic signals could be seen 
in large volume detectors as e.g.\ Superkamiokande. 

This kind of reasoning certainly applies to neutron 
invisible decay channels, e.g.\ $n \to 3\nu$.  
However, it is invalid for the $n-n'$ oscillation channel 
as far as the mirror particles $n'$, $p'$, $e'$
are exactly degenerate in mass with their ordinary partners $n$, $p$, $e$. 
Indeed, energy conservation allows the decay  
$(A,Z)\to (A-1,Z)+ n'$
(or $+ p'e'\tilde{\nu}'_e\,$)
only if the 
mass difference between the isotopes $(A,Z)$ and $(A-1,Z)$ 
is larger than $m_n$ (or $m_p + m_e$).
But then also the decay with neutron emission 
($p\, e\, \tilde{\nu}_e$ emission)
would be kinematically allowed 
and such a nucleus could not be stable.
One confirms by inspection that all nuclear ground states satisfying such conditions have short lifetimes rendering the extremely rare mirror channels invisible in 
practice~\cite{footnote3}.

Thus, the $n-n'$ oscillation cannot destabilize nuclei 
and so the only realistic limit remains  $\tau > 1$ s 
imposed by the data from the Grenoble experiment~\cite{Grenoble}. 

We discuss now the possible theoretical models for the operators 
(\ref{n-npr}) and their phenomenological implications. 
The contact terms of the form $\cO_9^{\rm mix}$ with $\cM \sim 10$ TeV,  
if valid at TeV energies, could have interesting consequences 
for future high energy colliders as the LHC.  
Namely, the mirror baryons could be produced in proton collisions, 
which would be seen as processes with baryon number 
violation and large missing energy. 

For example, in theories with large extra dimensions \cite{ADD}, 
the ordinary and mirror worlds 
can be conceived as parallel 3-dimensional branes   
embedded in a higher dimensional space, 
where O-particles with a gauge group $G$ are localized on one brane 
and M-particles with a gauge group $G'$ on another brane, 
while gravity propagates in the bulk. In the context of such a theory, 
operators $\cO_9^{\rm mix}$ with a cutoff $\cM\sim 10$ TeV   
could be induced as effects of the TeV scale quantum gravity. 
In addition, the baryon number can be related to a gauge symmetry 
in the bulk \cite{ADD}, in our case  $U(1)_{\bar B}$, $\bar{B}=B-B'$, 
that forbids the 
operators $\cO_9$, $\cO'_9$ leading to $n-\nbar$ oscillations.  

It is also possible that at TeV energies 
the terms (\ref{n-npr}) do not exist literally in the contact form,  
but are rather induced 
in the context of some renormalizable theory. 
Let us give a simple example of such a model. 
Assuming again $\bar{B}$ conservation, 
consider the Yukawa terms 
\be{S}
u d S + q q S + S^\ast d \cN +  
u' d' S' + q' q' S' + S^{\prime\ast} d' \cN'  
\ee 
where $S$ ($\bar{B}=-2/3$) is a color-triplet scalar $S$ with mass $M_S$,  
having precisely the same gauge quantum numbers 
as the right down-quark $d_{(R)}$,
and $S'$ ($\bar{B}=2/3$) is its mirror partner with the same mass $M_S$, 
whereas $\cN_{(R)}$ $(\bar{B}=-1)$ and $\cN'_{(R)}$ $(\bar{B}=1)$ 
are additional gauge singlet fermions, 
with a large mass term $\tilde{M} \cN \cN'$ 
(for simplicity, the Yukawa constants are assumed to be of order 1).  
Then, 
at energies $E \ll M_S,\tilde{M}$, 
the operators $\cO_9^{\rm mix}$ are induced 
with a cutoff scale 
$\cM \simeq (M_S^4 \tilde{M})^{1/5} $
(see Fig.\ \ref{fig1})
 \cite{footnote4}.

\begin{figure}[b]
\begin{center}
\includegraphics[width=60mm]{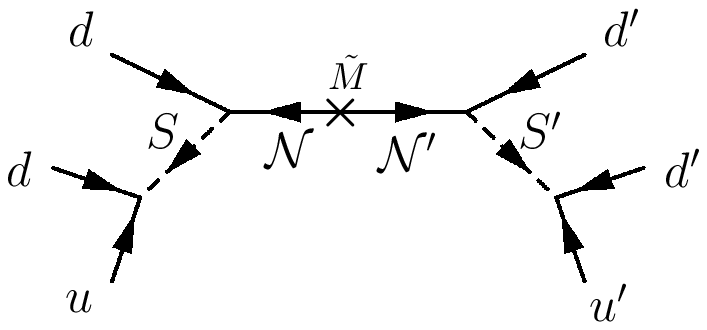}
\end{center}
\caption{\label{fig1} Diagram inducing 
neutron - mirror neutron mixing.}
\end{figure}

If e.g.\ both $M_S$ and $\tilde{M}$ are $\sim10$ TeV, 
then at TeV energies the operators (\ref{n-npr}) act in a contact form.
The scale
$\cM \sim 10$ TeV can be effectively obtained 
on other extremes, by taking e.g.\ 
$M_S\sim 200$ GeV and $\tilde{M} \sim  6 \times 10^{10}$ GeV, 
or  $M_S\sim 700$ GeV and  $\tilde{M} \sim 4\times 10^{8}$ GeV. 
In this case, while the detection of the scalar $S$ 
is within the reach of LHC, the mirror quarks cannot be produced 
as the messengers $\cN$, $\cN'$ 
are far too heavy. 

Let us address now the cosmological  limits. 
The most serious constraints come from the requirement that the
O- and M-worlds should have different temperatures  $T$ and $T'$
at the Big Bang Nucleosynthesis (BBN) epoch.  
The mirror sector would contribute to the universe expansion rate
as an effective number of extra neutrinos
$\Delta N_\nu \simeq 6.14\, (T'/T)^4$ \cite{BDM}, 
and thus  the bounds on $\Delta N_\nu$  demand  
that $T' < 0.5 \,T$ or so. 
This can be achieved by adopting the following paradigm \cite{IJMPA-B}:
at the end of inflation the O- and M-worlds are (re)heated in a 
non symmetric way, $T_r > T'_r$, which can naturally occur 
in the context of certain inflationary models;  
after (re)heating, 
$T < T_r$,
the processes between O- and M-particles are too slow  
to bring the two sectors in equilibrium, so that   
both systems expand adiabatically, 
maintaining the temperature asymmetry $T'<T$ in all subsequent epochs. 
  
The operators (\ref{n-npr}) induce 
collision processes as $udd \to $ $\bar{u}'\bar{d}'\bar{d}'$, etc.\
that cause entropy transfer between the ordinary and mirror sectors.  
Their effective rate scales as
$\Gamma = A \, T^{11}/\cM^{10}$,  
where the coefficient $A\sim 0.06$ 
accounts for phase space factors, 
and for $\cM \sim 10$ TeV they
would be in equilibrium 
at temperatures above $T_{\rm eq} \sim 0.5$ TeV or so. 
Thus, in order to fulfill the BBN request on the $T' < T$ asymmetry,   
the (re)heat temperature $T_r$ should be smaller than 
$0.5$ TeV~\cite{footnote5}. 

This applies only if at temperatures $T \sim T_{\rm eq}$ 
the operators (\ref{n-npr}) act in the contact form. 
In the context of the model presented above, 
that is true if the scalar $S$ is heavy enough, $M_S > 0.5$ TeV or so. 
For smaller $M_S $, however, 
the dominant process  at temperatures $T > M_S$  
would be rather $d \bar{S} \to \bar{d}'S'$.
Comparing  its rate  $\Gamma \simeq 10^{-2}\, T^3/\tilde{M}^2$ 
with the Hubble parameter 
$H = 1.7\, g_\ast^{1/2} T^2/\mpl$ 
($g_\ast \sim 10^2$), 
one obtains that these processes would bring the
two sectors in
equilibrium  only at temperatures above 
$T_{\rm eq} \sim $ $ 2\times 10^3 \, \tilde{M}^2/\mpl$, 
i.e., 
$T_{\rm eq} \sim $
$ 2\times 10^3 \,\cM^{10}/(M_S^8 \mpl) \sim $
$ 0.5 ~ {\rm TeV} \times (0.5~ {\rm TeV}/M_S)^8$. 

Hence, the output of the limit $T_r< T_{\rm eq}$ strongly depends on $M_S$. 
For $M_S < 0.5$ TeV it is milder:  
e.g.\ for $M_S \sim 200$ GeV   ($\tilde{M} \sim  6 \times 10^{10}$ GeV)   
it turns $T_r <  6\times 10^{5}$ GeV.   
But it sharply strengthens with increasing $M_S$, and for 
$M_S > 0.5$ TeV becomes roughly $T_r < 0.5$ TeV. 
This restricts the possible inflation scenarios  
and excludes many popular scenarios for primordial baryogenesis.

But on the other hand, the same
particle exchange processes  between O- and M-sectors, 
$udd \to \bar{u}'\bar{d}'\bar{d}'$ or $d \tilde{S} \to \tilde{d}'{S}'$, 
once they are out of equilibrium, violate $B$, $B'$ and possibly also $CP$, 
could provide a plausible low-temperature baryogenesis mechanism 
for both observable matter (O-baryons) and dark matter (M-baryons), 
 along the lines of the leptogenesis scheme via the  scatterings 
 $l\phi \to \bar{l}' \bar{\phi}'$  suggested in~\cite{BB-PRL,IJMPA-B}. 

The fast $n-n'$ oscillation could have intriguing implications 
for ultra high energy (UHE) cosmic rays.  
Namely, when a UHE proton $p$  
scatters a relic photon, it produces a neutron $n$ that promptly oscillates   
into a mirror neutron $n'$ which then decays into a mirror proton $p'$.
The latter undergoes a symmetric process, scattering a mirror relic photon 
and producing back an ordinary nucleon. 
However, as the M-sector is cooler, $T'/T < 0.5$, 
the mean free path of $p'$ is larger by a factor of $(T/T')^{3}$
than that of ordinary protons ($\sim 5$ Mpc). 
In this way, the UHE protons could propagate 
at large cosmological distances without significant energy losses.  
This may relax the Greisen-Zatsepin-Kuzmin cutoff in the cosmic ray spectrum \cite{GZK}
and also explain the correlation between the observed UHE protons 
and far distant sources as BL Lacertae~\cite{TT}. 

Concluding, we observed an intriguing loophole in the 
physics of such a familiar and long studied particle  
as the neutron: the existing experimental data do not exclude 
that its oscillation time into a mirror partner may be as small as 1 s. 
This oscillation, however, is impossible for neutrons bound in nuclei,   
while for free neutrons it is suppressed by matter and magnetic field 
effects~\cite{footnote6}. 

Our suggestion is falsifiable at small costs. 
"Table-top" experiments searching for neutron disappearance ($n\to n'$) 
and regeneration ($n\to n' \to n$), 
performed in proper "background-free" conditions, 
may discover the neutron - mirror neutron oscillation 
and thus reveal the existence of sterile partners of nucleons,   
opening up a window to the mirror world with a number of 
serious implications in astrophysics and cosmology.

\vspace{5pt}


We are grateful to V. Berezinsky, S.L. Glashow, 
S.\ Gninenko, Y.\ Kamyshkov, V.\ Kuzmin, R.N.\ Mohapatra, 
V.\ Nesvizhevsky, T.\ Soldner and I.\ Tkachev for useful discussions.
The work of Z.B.\ is partially supported by the 
MIUR biennal Grant No.\ PRIN/2004024710/002, 
and that of L.B.\ 
by the Grant POCTI/FNU/43666/2002.


\vspace{-10pt}



\begin{thebibliography}{99}
\bibitem{footnote1}{For simplicity, we omit the gauge, spinor and family indices. 
In the following, whenever there is no ambiguity, 
the chiral symbols L and R are omitted as well. 
 Antiparticles will be marked as $\tilde{p}, \tilde{n}$, etc..}
\bibitem{footnote2}
{If the Majorana  operators $\cO_5,\cO'_5$ are also forbidden 
by $\bar{L}=$ $  L - L'$ conservation, then 
$\cO^{\rm mix}_5$ 
yields light Dirac neutrinos with masses 
$m_{\nu} \sim \langle\phi\rangle^2/M$,  
where the right-handed components are mirror anti-neutrinos: 
$\nu'_R = {{\nu}'_L}^c$. 
}
\bibitem{footnote3}
{In principle, the $n-n'$ mixing could destabilize also
objects like neutron stars. 
However, the effective mixing angle is suppressed by the neutron 
Fermi energy of several MeV, $\theta^2_{\rm eff} < 10^{-44}$, 
and one can easily estimate that the conversion time 
into a mirror neutron star exceeds by many orders of magnitude 
the age of the Universe.
}
\bibitem{footnote4}
{The $n-\nbar$ oscillation can be also accommodated 
by including $\bar{B}$ violating mass terms 
$\mu (\cN\cN + \cN' \cN')$
with $\mu \ll \tilde{M}$, that give rise to 
the operators $\cO_9$, $\cO'_9$  cutoff by 
$\cM^5_{n\nbar} \sim (\tilde{M}/\mu)\cM^5$.    
Then, for $\cM \sim 10$ TeV, the bound $\dm_{n\nbar}< 10^{-23}$ eV 
requires $\mu/\tilde{M} < 10^{-8} $.
}
\bibitem{footnote5}
{The $n-n'$ oscillation itself has no effect at the BBN epoch 
because it is strongly suppressed by matter effects. 
}
\bibitem{footnote6}
{The $n-n'$ and $\nu-\nu'$ mixings induce together 
an effective operator $\sim G_F^2 m_\nu \dm \, (pe)(p'e')$ 
that mixes
the hydrogen atom H and its mirror atom H$'$.
A larger effect may result directly from 
independent operators $\propto (qque)(q'q'u'e')$, etc.. 
The limits on the H-H$'$ oscillation, also sensitive to 
matter and magnetic field effects, will be discussed elsewhere. 
}
\end{thebibliography}
\end{document}